\documentstyle[twoside,fleqn,espcrc2]{article}
\def\prb{Phys. Rev. B}
\def\prl{Phys. Rev. Lett.}

\def\be{\begin{equation}}
\def\ee{\end{equation}}
\def\ba{\begin{eqnarray}}
\def\ea{\end{eqnarray}}

\def\YBCO{YBa$_2$Cu$_3$O$_{7-\delta}$}

\def\BSCCO{Bi$_2$Sr$_2$CaCu$_2$O$_{8+\delta}$}
\def\C60{A$_x$C$_{60}$}
\def\LNSCO{La$_{1.6-x}$Nd$_{0.4}$Sr$_x$CuO$_{4}$}

\def\hty{high temperature superconductivity}
\def\hts{high temperature superconductors}

% ...

\newcommand{\AmS}{{\protect\the\textfont2
 A\kern-.1667em\lower.5ex\hbox{M}\kern-.125emS}}

\title{Pairing and Phase Coherence in High Temperature Superconductors}
\author{V.~J.~Emery,
\address{Dept. of Physics,
Brookhaven National Laboratory,
Upton, NY  11973}
\thanks{Supported by the Division of Materials Science,
U. S. Department of Energy under contract No. DE-AC02-76CH00016.}
S.~A.~Kivelson, and O.~Zachar
\address{Dept. of Physics, University of California at Los Angeles,
Los Angeles, CA 90095}
\thanks{Supported in part by the National Science Foundation 
grant number DMR93-12606.}}

\begin{document}

\begin{abstract}

Mobile holes in an antiferromagnetic insulator form a slowly 
fluctuating array of quasi one-dimensional metallic 
stripes, which induce a spin gap or pseudogap in the intervening 
Mott-insulating regions. 
The mobile holes on an individual stripe acquire a 
spin gap via pair hopping between the stripe and its environment; {\it i.e.} 
via a magnetic analog of the usual superconducting proximity effect.
This process is the analog of pairing in conventional superconductors.
At non-vanishing stripe densities, Josephson coupling between stripes produces 
a dimensional crossover to a state with long-range superconducting phase 
coherence. In contrast to conventional superconductors, the superconducting 
state is characterised by a high density of (spin) pairs, but the phase 
stiffness, which is determined by the density and mobility of holes on
the stripes, is very low.

\end{abstract}

\maketitle

\section{Introduction}

Superconductivity in metals requires pairing and long-range phase coherence
\cite{bcs}. In clean homogeneous conventional superconductors, pairing 
involves a relatively small fraction of the conduction electrons \cite{conden}, 
but the superfluid density (which determines the phase stiffness)
involves all of them. 
Here we argue that, in the {\hts} \cite{bm}, it is just the other way round: 
most of the holes are involved in pairing, but the superfluid density is
proportional to the density of doped holes.

The poor phase stiffness of the {\hts} is well known, and it implies that
the transition temperature $T_c$ is much less than the pairing temperature, 
$\sim \Delta_0 /2$, where $\Delta_0$ is the energy gap at zero temperature. 
This separation of the energy scales 
has been demonstrated on phenomenological grounds \cite{nature}. 
In underdoped and optimally doped materials, $T_c$ is determined by 
the onset of phase coherence, which occurs at a temperature of 
about ${\hbar}^2 n_s(0)/4m^*$. Here $n_s(0)$, the two-dimensional 
superfluid density at zero temperature, determines the phase stiffness, and 
it is low because the {\hts} are doped Mott insulators.

The need for a different approach to pairing is clear in view of the 
difficulty of achieving a high transition temperature via the
conventional mechanisms. The problem is that a high pairing scale requires a 
strong attractive interaction which may favor other instabilities or,
alternatively, produce a large mass renormalization which depresses the phase 
coherence temperature.
Moreover, as the pairing energy increases, retardation becomes less
effective, so it is all the more difficult to overcome the Coulomb repulsion. 
This problem is especially acute for the {\hts}, which are 
doped Mott insulators with relatively poor screening. Angle resolved 
photoemission spectroscopy \cite{arpes} (ARPES) suggests that the energy gap
has the form $\cos k_x - \cos k_y$. This implies that, in real space, the gap 
function (and hence the pairing force) has a range of one lattice 
spacing, where the bare Coulomb interaction is very large. 

In short, a theory of {\hty} must show how to obtain a large temperature 
scale for local superconductivity, without detriment to global phase 
coherence, despite poor screening of the Coulomb interaction. There are
several phenomenological constraints on such a theory. First of all the order 
parameter has charge $2e$ \cite{2e}, {\it i.e.} there is some kind of pairing.
However ARPES experiments, which show that the chemical potential is 
near the center of the bare hole band, rule out real-space pairing, which 
anyway is implausible for a $d$-wave superconductor with a 
strong and poorly-screened Coulomb repulsion between electrons.
More generally, the conventional view of superconductivity as a Fermi surface 
instability resulting from an attractive interaction between quasiparticles 
is inapplicable, since, according to analyses of resistivity \cite{badmetal} 
and ARPES data \cite{noqp} there are no well-defined quasiparticles or 
Fermi surface in the normal state of {\hts}.

\section{Spin-gap Proximity effect}

We have proposed \cite{ekzsg} that the {\hts} satisfy these constraints
in a unique manner. The mechanism of pairing is a form of internal magnetic 
proximity effect in which a spin gap is generated in {\it Mott-insulating} 
antiferromagnetic (AF) regions through spatial confinement by charge stripes, 
and communicated to the stripes by pair hopping. Many of the problems listed
above are avoided because the order parameter relies on spin-pairing and does 
not require the existence of {\it bound} pairs of holes. The successive steps 
in the argument are as follows:

An AF insulator tends to expel holes
\cite{spbag}. For neutral holes this leads to phase separation into hole-rich 
and hole-free regions \cite{ekl}, whereas, for charged holes, it gives rise
to {\it local} charge inhomogeneity, typically in the form of topological 
doping \cite{topo}, in which metallic stripes are separated by 
insulating antiphase AF regions  \cite{ute}.  
There is much experimental evidence of ordered or fluctuating 
structures of this kind \cite{jtbj,zaanen}. 

Topological doping \cite{topo} is a general feature of doped Mott 
insulators, and it amounts to a strong (anti)correlation of spin and charge. 
However, within a metallic stripe or the intervening undoped regions there is a
separation of spin and charge \cite{LE}, as in the one-dimensional electron 
gas (1DEG). In the 1DEG, the boson representation of the operator 
$\psi_{1,\uparrow} \psi_{2,\downarrow} - 
\psi_{1,\downarrow} \psi_{2,\uparrow}$ (which annihilates pairs of fermions
at opposite Fermi points, labelled by 1,2) may be written in the form
$\exp(- i \theta_c) \cos \phi_s$, where $\theta_c$ and $\phi_s$ are associated 
with the charge and spin degrees of freedom respectively. This is an 
{\it operator relation}, and $\theta_c$, the superconducting phase field,
is a property of the charge modes. On the other hand, pairing (or, 
equivalently a well-defined amplitude) corresponds to a finite expectation 
value of $\cos \phi_s$, which requires a spin gap \cite{strong}.

 A large spin gap (or pseudogap) arises naturally in a spatially-confined
AF region, such as the medium between stripes. This behavior is 
well documented for clusters \cite{lintang}, frustrated spin chains \cite{mg}, 
and spin ladders \cite{ladder}. In general, if there are $N$ spins in a unit 
cell, all spin excitations are gapped if $N$ is even, but one excitation is 
ungapped if $N$ is odd, {\it i.e.} there is a pseudogap.  
Such a spin gap does not conflict with the Coulomb interaction
since the energetic cost of having localized holes in Cu $3d$ 
orbitals has been paid in the formation of the material. 

The spin degrees of freedom of the 1DEG acquire a spin gap by pair 
hopping between the stripe and the AF environment \cite{ekzsg}.
Because of the {\it local} separation of spin and charge,
the spin-gap fixed point is stable even in the presence of strong Coulomb
interactions, and there is no mass renormalization to depress the onset of 
phase coherence. Thus the phase stiffness of the mobile holes on the stripes 
is large enough to give a high superconducting transition temperature.

Both superconducting and charge density wave correlations 
develop on a given stripe. They compete at longer length scales,
although they may coexist in certain regions of the phase diagram.

\subsection{Symmetry of the order parameter:} 
If stripe order breaks the four-fold 
rotational symmetry of the crystal, the superconducting order will have
strongly mixed extended-$s$ and $d_{x^2-y^2}$ symmetry.  This
will happen in a stripe-ordered phase, such as in {\LNSCO}, or in a possible 
``stripe nematic'' phase, in which the stripe positional order is destroyed 
by quantum or thermal 
melting or quenched disorder, but the stripe orientational order is preserved.
Such  phases also would display large induced asymmetries in the
electronic response in the $ab$-plane.

In tetragonal materials, the order 
parameter must have a pure symmetry, but the way in which it emerges from the
short-distance physics is very different from more conventional routes.
If doping is not too high, $d_{x^2-y^2}$ order
should give the long distance behavior because the extended-$s$ 
order parameter ($\cos k_x + \cos k_y$) is small on the Fermi surface of the 
noninteracting system. Nevertheless,
if the stripe fluctuations exhibit substantial orientational order 
at intermediate length scales, the interplay between the two
types of superconducting order may be more 
subtle than in conventional, homogeneous materials. 

\subsection{Phase diagram}

A feature of the stripe model is that, in lightly doped
materials, the temperature scale, 
$T_{pair}$, at which pairing occurs on a single stripe is parametrically
larger than the superconducting transition temperature, $T_c$, which is
governed by the Josephson coupling between stripes.  Moreover, 
both $T_{pair}$ and $T_c$ must be less than the temperature scale, $T_{AF}$, 
at which the {\it local} AF correlations are developed.
The crossovers observed experimentally in
underdoped high temperature superconductors have tentatively been 
identified \cite{batloggemery,ekzsg} with these two phenomena.

\section{Evidence for Spin Pairing}

The idea that there is pairing in a large range 
of temperatures above $T_c$ is an immediate consequence 
of the fact that $T_c$ is significantly suppressed by phase fluctuations; 
{\it i.e.} that $T_c$ is close to the value obtained from  
the phase stiffness derived from
the London penetration depth at zero temperature. 
Moreover the idea that ``pairing''  
means a high density of singlets between stripes, rather than {\it bound}
Cooper pairs provides a very natural explanation
of the spin-gap behavior that has been widely observed in planar copper
NMR measurements in underdoped cuprates \cite{nmrgap}.  
The interpretation of the spin gap
as a superconducting gap has recently received considerable support from 
ARPES experiments \cite{shen} which find that the magnitude and wave 
vector dependence of the pseudogap above $T_c$ is similar to that of the
gap seen well below $T_c$  in both underdoped and optimally doped materials.
The temperature above which this gap structure disappears correlates
well with the pairing scale deduced from NMR.
The in-plane optical response, which shows a narrowing of the ``Drude-like''
peak, rather than a pseudogap structure \cite{optical}, also is consistent
with the idea of spin pairing.

\subsection{Spectroscopic evidence}

The most direct evidence for the proposed pairing 
mechanism would be the observation of an isolated spin-1, charge 
zero excitation, with an energy of order the superconducting gap,
which could be identified as the low-energy excitation of a small 
undoped region \cite{physica}. Neutron scattering experiments 
on optimally-doped {\YBCO} \cite{keimer} have indeed found a spin triplet
excitation, with wave vector $(\pi/a,\pi/a)$ and energy 40meV which
first appears in the neighborhood of $T_c$. At lower doping \cite{under}, the 
energy of the mode decreases, and it first appears above $T_c$, although its 
intensity is enhanced below $T_c$. Recently, ARPES data in {\BSCCO} have
been interpreted as evidence for such a mode \cite{norman}.

It is natural to interpret these experiments in terms of excitations of the 
pairs that give rise to {\hty} \cite{keimer}. However it is important 
to note that the observed cross section and Cu-Cu bilayer modulation
of the triplet mode are very close to those
of AF spin waves in the undoped antiferromagnet. 
(See {\it e.g.} Fig. 4 of ref. \cite{keimer}.) This strongly suggests that the mode is 
associated with undoped regions of the material and not with mobile holes.
The point is that the wave function of the mobile holes has a strong 
admixture of O(2p) orbitals and therefore a different magnetic form factor
than in the undoped material, in which Cu(3d) orbitals predominate.
Furthermore, in a conventional BCS picture, the intensity of the peak 
would be quite small because it is proportional to $N(0) \Delta_0$.

In our model \cite{ekzsg}, there is a composite order (involving the 
stripe and the AF regions), which does not break gauge invariance, but 
associates a superconducting phase factor $\exp(- i \theta_c)$
with singlets in the AF regions. Thus, the triplet mode should appear below
$T_{AF}$, and its intensity should begin to increase below $T_{pair}$ 
(where significant phase fluctuations begin),
then grow substantially near $T_c$ (where long-range phase order is 
established). In optimally-doped {\YBCO}, $T_{AF}$, $T_{pair}$ and $T_c$ are 
very close together \cite{batloggemery,ekzsg}, so it is reasonable that the 
mode is first observed close to $T_c$.
The position and width of the peak reflect the wave function
of the singlets; it occurs at $(\pi/a, \pi/a)$ because the singlets involve
spins on opposite sublattices, and it it corresponds to a singlet size of a 
few lattice spacings. 
A more detailed description of this behavior will be 
given in a future publication.


\begin{thebibliography}{9}
\bibitem{conden} The density of electrons involved in 
particle-hole mixing is $N(0)\Delta_0$, where
$2N(0)$ is the density of states at the Fermi level.
\end{thebibliography}
\end{document}